\documentclass[preprint,aps,prc,a4paper,superscriptaddress,showpacs,amsmath,amssymb]{revtex4-1}
\usepackage{amssymb}
\usepackage{amsmath}
\usepackage{graphicx}
\usepackage{booktabs,longtable}
\usepackage{multirow}
\usepackage{color}
\usepackage{CJK}

\begin{document}
\begin{CJK*}{GBK}{kai}

\preprint{API/123-QED}

\title{Valence Neutron-Proton Orientation in Atomic Nuclei}

\author{J.~G.~Wang}
\affiliation{Key Laboratory of High Precision Nuclear Spectroscopy
and Center for Nuclear Matter Science, Institute of Modern
Physics, Chinese Academy of Sciences, Lanzhou 730000, People's
Republic of China}
\affiliation{School of Nuclear Science and Technology, University of Chinese Academy of Science, Beijing
100049, People's Republic of China}
\author{M.~L.~Liu}\email{liuml@impcas.ac.cn}
\affiliation{Key Laboratory of High Precision Nuclear Spectroscopy
and Center for Nuclear Matter Science, Institute of Modern
Physics, Chinese Academy of Sciences, Lanzhou 730000, People's
Republic of China}
\affiliation{School of Nuclear Science and
Technology, University of Chinese Academy of Science, Beijing
100049, People's Republic of China}

\author{C.~M.~Petrache}
\affiliation{Centre de Spectrom\'{e}trie Nucl\'{e}aire et de
Spectrom\'{e}trie de Masse, Universit\'{e} Paris-Sud and
CNRS/IN2P3, B\^atiment 104-108, Orsay 91405, France}

\author{K.~K.~Zheng }
\affiliation{Key Laboratory of High Precision Nuclear Spectroscopy
and Center for Nuclear Matter Science, Institute of Modern
Physics, Chinese Academy of Sciences, Lanzhou 730000, People's
Republic of China}
\affiliation{School of Nuclear Science and
Technology, University of Chinese Academy of Science, Beijing
100049, People's Republic of China}
\affiliation{Centre de
Spectrom\'{e}trie Nucl\'{e}aire et de Spectrom\'{e}trie de Masse,
Universit\'{e} Paris-Sud and CNRS/IN2P3, B\^atiment 104-108, Orsay
91405, France}

\author{X.~H.~Zhou }
\affiliation{Key Laboratory of High Precision Nuclear Spectroscopy
and Center for Nuclear Matter Science, Institute of Modern
Physics, Chinese Academy of Sciences, Lanzhou 730000, People's
Republic of China}
\affiliation{School of Nuclear Science and
Technology, University of Chinese Academy of Science, Beijing
100049, People's Republic of China}
\author{Y.~H.~Zhang}
\affiliation{Key Laboratory of High Precision Nuclear Spectroscopy
and Center for Nuclear Matter Science, Institute of Modern
Physics, Chinese Academy of Sciences, Lanzhou 730000, People's
Republic of China}
\affiliation{School of Nuclear Science and
Technology, University of Chinese Academy of Science, Beijing
100049, People's Republic of China}

\date{\today}

\begin{abstract}
It is shown that the renormalized nuclear deformations in
different mass regions can be globally scaled by two probability
partition factors of Boltzmann-like distribution, which are
derived from the competing valence $np$ and like-nucleon
interactions. The partition factors are simply related to the
probabilities of anti-parallel and fully-aligned orientations of
the angular momenta of the neutrons and protons in the valence
$np$ pairs, responsible for spherical- and deformed-shape phases,
respectively. The partition factors derived from the renormalized
deformations are also present in the new scaling law for the
energies of the first $2^+$ states. A striking concordance between
the distributions of the renormalized deformations and of the
newly introduced parameter for the energies of the first $2^+$
states over the extended mass region from Ge to Cf is achieved,
giving strong support to the existence of two phases: anti-aligned
and fully-aligned subsets of $np$ pairs.

\end{abstract}

\pacs{21.10.Re, 23.20.Lv, 27.60.+j, 25.70.Jj}

\maketitle

\end{CJK*}

The atomic nucleus is a binary-fermion quantum many-body system.
Valence neutron-proton ($np$) and like-nucleon ($nn$ and $pp$)
correlations are two fundamental ingredients in determining
nuclear structure \cite{Cejnar2010}. Bohr, Mottelson and Pines
pointed out the like-nucleon pairing in the ground states of
even-even nuclei \cite{Bohr1958}, while one may expect the extreme
(parallel or anti-parallel) coupling of neutron and proton angular
momenta in nuclei, which is induced by the strongest residual
interactions \cite{Schiffer1976,Paar1979,Heyde1985}. Thus, there
should be a strong competition energy-wise between $np$ pairs
having parallel and anti-parallel coupling of the angular momenta
of the involved nucleons \cite{Macchiavell2000,Frauendorf2014}. It
has been long recognized that the $np$ residual interaction is
primarily responsible for configuration mixing of valence nucleons
in open shells and therefore plays a key role in the onset of
nuclear deformation and collectivity \cite{Shalit1953,Talmi1962}.
A parameter $P=N_n N_p/(N_n+N_p)$ has been introduced to express
the average number of interactions of each valence nucleon with
those of the other type, which can be related to the relative
integrated strengths of $np$ and like-nucleon pairing interactions
\cite{Casten1987}. $N_n$ and $N_p$ refer to the numbers of
particles or holes relative to the closest shell closures
 \cite{Casten1985-1,Casten1985-2}. The $P$ parameter provides an universal
description of the nuclear structure in heavy nuclei and its
evolution from spherical single-particle to deformed collective
motion induced by the single-particle coherent contributions of
the nucleons when the Fermi surfaces are away from the shell
closures \cite{Mayer1948,Mayer1950,Haxel1949,Bohr1969, Bohr1975}.
 $P$ is also the controlling
parameter of the shape phase transitions in nuclei \cite{Cejnar2010, Subir1999}. It is worth noting that
the nucleus, being a finite mesoscopic system, exhibits gradual
phase transitions, unlike the infinite systems for which the changes are sharp.

As far as the anti-parallel coupling of the neutron angular
momentum $j_n$ and proton angular momentum $j_p$ is concerned, the
$np$ interaction between non-identical orbits is shared by the
T~=~0 isospin-scalar and T~=~1 isospin-vector channels. In modern
nuclear physics, special attention has been paid to the T~=~1 $np$
pairing of nucleons occupying identical orbits
\cite{Macchiavell2000,Frauendorf2014}. The multiplet subset with
anti-parallel coupling corresponds to the smallest number $ 2\mid
j_n-j_p\mid +1$ of $np$ interactions, while the largest number $
2(j_n+j_p)+1$ arises from the parallel multiplet subset. The total
number of available $np$ parallel interactions obtained by summing
over all single-particle orbits in the open shells, is generally
much larger than that of total $np$ anti-parallel interactions.
For $sd$ shells, the former (latter) amounts to 63 (25). In the
large $N_n N_p$ limit, if one assume nearly-degenerate or
comparable interaction energies for the configurations of parallel
and anti-parallel coupling, it results in that the majority of
$np$ interactions are contained in the fully-aligned subsets.

It is well known that a $np$ pair in identical $j$ orbits can
couple to angular momentum 1 or $2j$, and isospin $T=0$
\cite{Schiffer1976}, satisfying thus the antisymmetry of the total
wave function. The total spin in the fully-aligned configuration
is $S=1$. For neutron and proton occupying non-identical orbits,
the fully-aligned configuration is different. The single-particle
orbits can be divided into two categories depending on the
spin-orbit $ls$ coupling. Let $j_>=l+1/2$ and $j_<=l-1/2$ denote
the angular momenta resulting from parallel and anti-parallel $ls$
couplings, respectively. If both neutron and proton occupy the
non-identical $j_>$ orbits, the spin of the fully-aligned
configuration is $S=1$. The antisymmetry of the total $np$ wave
function leads to a relatively pure T~=~0 $np$ pair. Here we adopt
the principle that the central force dominates the nuclear force
and requires the spatial symmetry. For the other combinations, the
fully-aligned state includes T~=~0 and T~=~1 components. Taking
into account that the majority of $np$ interactions come from the
fully-aligned states and the T~=~0 interaction is stronger, we can
therefore infer the existence of a T~=~0 condensate of $np$ pairs
in the large $N_n N_p$ limit. However, the probability partition
between the two kinds of extreme coupling has so far been missing.
In this Letter, we present for the first time the probabilities of
anti-aligned and fully-aligned subsets of $np$ pairs in nuclei.

It is the fully-aligned subset of $np$ pairs that plays the most
important role in the onset of deformation and collective
rotational motion, as discussed earlier in Refs.
\cite{Mulhall1964,Sips1966}, but without mentioning the $np$
pairs. Danos and Gillet therefore introduced a concept of two
identical chains consisting of fully-aligned $np$ pairs for $N=Z$
even-even nuclei to interpret the collective rotation as the
gradual alignment of the paired chains \cite{Danos1966,Danos1967}.
They are called Ising-like chains hereafter because the $np$
angular momentum is an analogue of the Ising spin residing on the
Ising chain \cite{Ising1925,Subir1999}. The absence of deformation
and anti-parallel $np$ pairs in the Ising-like chains discussed by
Danos and Gillet \cite{Danos1966,Danos1967} for $N=Z$ nuclei,
makes the chains imperfect and unrealistic when applied to real
nuclei. We extend here the Ising concept to $N \neq Z$ cases, also
including the anti-parallel $np$ pairs. An even-even nucleus is
composed of $N_n/2$ valence-neutron and $N_p/2$ valence-proton
pairs. $N_n/2$ neutrons and $N_p/2$ protons can be picked out from
the like-nucleon pairs to form $\sqrt{N_nN_p}/2$ effective $np$
pairs despite the fact that non-integer values are obtained for
unequal numbers of valence neutrons and protons. The $np$ pairs
with two kinds of extreme (anti-parallel and fully-aligned)
coupling are the building blocks of a macroscopic chain carrying a
spin $J_{np}$ which can vary between a minimum ($J_{np}^{min}$)
and maximum ($J_{np}^{max}$), following an unknown distribution
with some probability. The remaining nucleons not involved in the
construction of Ising-like chain, constitute the other macroscopic
twin chain having opposite spin. The two Ising-like chains are
paired in the ground state and the spins of the constituents
gradually align along the rotation axis for rotation motion.

The total number of effective $np$
pairs which is $2\sqrt{N_nN_p}/2=\sqrt{N_nN_p}$ for two Ising-like chains, results from the number $N_nN_p$
of $np$ pairs formed by $N_n$ neutrons and $N_p$ protons and
the synthesis probability of a $np$ pair. Since the probability
of selection of a neutron and a proton from $N_n/2$ pairs of
neutrons and $N_p/2$ pairs of protons is $1/N_n$ and $1/N_p$,
respectively, the synthesis probability of a $np$ pair is
defined as the overlap of the wave functions of a neutron and a
proton \cite{Arrington2012,Hen2017}, normalized to the respective
square root of selection probabilities $\sqrt{1/N_n}$ and
$\sqrt{1/N_p}$, which is $1/\sqrt{N_nN_p}$. One can verify that
for $ N_n=N_p$, the number $\sqrt{N_nN_p}$ of effective $np$
pairs becomes $ N_n$ or $ N_p$, which shows that the proposed
extension of the Ising concept to $N \neq Z$ cases is coherent.

One generally believe that the configuration mixing is responsible
for deformation \cite{Talmi1962}. It is certainly true for the
Ising-like chains. The mixing is achieved in a series of chains
with spins $J_{np}\otimes J_{np} = J\otimes J,(J-1)\otimes(J-1),
(J-2)\otimes(J-2),\cdots$, where $2J$ represents the nuclear total
angular momentum. The mixing mechanism is similar to that
generated by the quadrupole field $r^2Y_{20}\sim z^2-x^2\sim
r(z-x)$ acting on a single-particle, which mixes the unique-parity
Nilsson
 orbits \cite{Nilsson1955}. As for the
macroscopic chains, the difference $R_z-R_x$
contained in the whole quadrupole operator $R^{2}Y_{20}$ is
effective in inducing the mixing, which accounts for the
deformation mechanism of Ising-like chains in the framework of
the spherical shell model \cite{Talmi1962}. The Ising-like chains
therefore bridge the spherical and deformed shell models
\cite{Nilsson1955,Caurier2005,Zuker1995,Elliott1958-1,Elliott1958-2,Iachello1987}.
The axially symmetric quadrupole deformation is related to the
asymmetry of the long and short axes through the parameter $\beta\sim(R_z-R_x)/(1.2A^{1/3})$, which is positive for
prolate ($R_z>R_x$) and negative for oblate ($R_z<R_x$) shapes. We realize that
the renormalized shape parameter $\beta A^{1/3}$ is a measure of the extent of effective
quadrupole mixing of Ising-like chains and might be considered as a global
observable over the entire nuclear mass table.

Much effort has been devoted to the regional systematics of the
quadrupole deformation parameter $\beta$ of the even-even isotopes
from Germanium to Californium, the regions being divided by closed
shells or sub-shells \cite
{Raman1988,Cottle1991,Margraf1992,Zhao2000,Pritychenko2016}. The
present study, which employs the renormalized deformation $\beta
A^{1/3}$, goes beyond the systematics over limited mass regions.
We employ the correlation scheme of the renormalized deformation
$\beta A^{1/3}$ and $P$ factor to examine the independence and
judge the adequacy of the description employing Ising-like chains.
All the data are taken from the recently compiled Table~3 in
Ref.~\cite {Pritychenko2016}.

\begin{figure}[bt]
\includegraphics[width=0.47\textwidth]{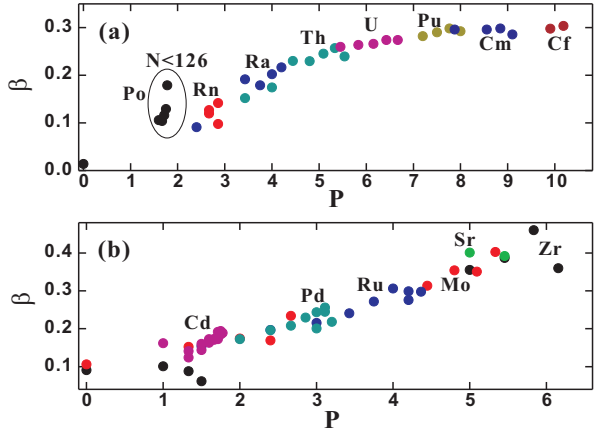}
\caption{(Color online) $\beta$ \emph{vs} $P$ for (a)
Polonium-Californium isotopes; (b) Strontium-Cadmium isotopes with
$N>50$. Note that the enclosed Po isotopes deviate from the
systematics because of their neutron numbers less than the magic
number 126. Zr-Mo (Sr) isotopes undergo a sharp change of shape at
$N=60$ due to the collapse of proton sub-shell closure 38 and the
creation of sufficient valence space for deformation and the closed shell $%
Z=50~(28)$ has to be used as the new reference core. The deviation
of $\beta$ values of $^{92-96}$Zr is due to the proton sub-shell
closure 40 resulting in the effective $N_p=0$ rather than 2
relative to 38, which are omitted in the following figures. }
\label{fig1}
\end{figure}

Nuclear deformations in different mass regions always show a rise
until saturation for nuclei with mass number $A>150$, after
undergoing an inflection at a certain $P_c$ value. Two typical
examples are shown in Fig.~\ref{fig1}, for Strontium-Cadmium and
Polonium-Californium isotopes. The overall growth of the
deformation can be easily understood as induced by the enlarged
$np$ interaction energy which increases with increasing $ N_nN_p$,
while the saturation can be attributed to the attenuation of the
interaction between valence neutrons and protons near mid-shells
occupying extremely different orbits \cite {Casten1988,Zhang1989}.
However, it is obviously questionable whether the saturation
exists in any light-mass region. Comparing the resemblance or
difference of such plots with those of a complete set covering
several regions would contribute to solve the problem.

\begin{figure}[tb]
\includegraphics[width=0.40\textwidth]{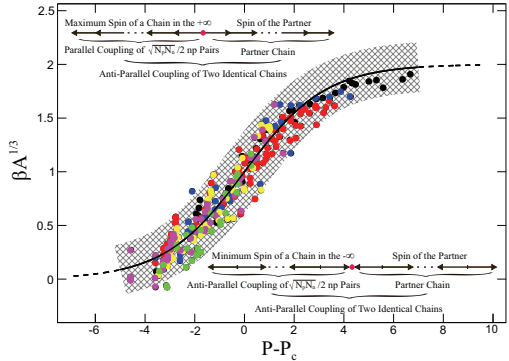}
\caption{(Color online) Renormalized $\beta A^{1/3}$ \emph{vs}
$P-P_c$ for the ground state of even-even nuclei: Po-Cf (black),
Dy-Hg (red), Te-Gd with $N>82$ (blue), Te-Gd with $ N<82$
(yellow), Kr-Cd with $N>50$ (pink), and Ge-Zr with $N<50$ (green),
where $P_c$ is taken from Table~\ref{tab1}. Ce-Dy (Er-Hf) isotopes
undergo a sharp change in the shape and excitation energy at
$N=88$ due to the collapse of proton sub-shell closure 64 and the
closed shell $Z=50$ (82) has to be used as the new reference core.
We propose two identical Ising-like chains consisting of parallel
or fully-aligned $np$ pairs for simplicity responsible for the
unification of nuclear spherical single-particle and deformed
collective properties as $P-P_c$ going from the negative to
positive infinite limits corresponding to the infinite limits of
valence-nucleon numbers.} \label{fig2}
\end{figure}

\begin{table}[b]
\caption{Inflection $P_c$ values of the $P$ parameter and saturated energy ratios $R_s$ for
given mass regions.} \label{tab1}
\begin{ruledtabular}
\begin{tabular}{cccccccc}
\multirow{2}*{Region} & \multirow{2}*{Po-Cf} & \multirow{2}*{Dy-Hg} & Te-Gd & Te-Gd  &  Kr-Cd & Ge-Zr \\
 & & & ($N$$>$82) & ($N$$<$82) & ($N$$>$50) & ($N$$<$50) \\
\hline
$P_{c}$ & 3.53 & 4.6 & 3.6 & 4.45 & 4.6 & 4.75 \\
$R_{s}$ & 32.78 & 16.00 & 17.16 & 12.94 & 12.70 & 10.86 \\
\end{tabular}
\end{ruledtabular}
\end{table}

It is well known that the nuclear deformations show an overall
decrease from the lightest to heaviest nuclei
\cite{Pritychenko2016}. The initial $\beta$ value of $A\sim$~110
nuclei tends to be 0.07 larger than that of $A\sim$~220 nuclei,
while one expects an increase of $\beta$ by a factor of near two
for $P=6$ (see Fig. \ref{fig1}). Taking into account the
dependence of $\beta$ on the asymmetry of the nuclear radius
$\beta\sim(R_z-R_x)/(1.2A^{1/3})$, the decrease of the deformation
$\beta$ with increasing $A$ should be associated with the
$A^{-1/3}$ dependence \cite{Pritychenko2016}. We found that the
renormalized deformations, independent of the mass region, are
universally valid around their inflection $P_c$ parameters given
in Table~\ref{tab1}. A similar universality is also present  for
the excitation energies, around the same $P_c$ values (see the
following discussion). Figure~\ref{fig2} shows that the
renormalized deformations of nuclei in different mass regions fall
into a twisted band centered around the $P-P_c=0$ inflection
point. A striking regularity is therefore achieved. A curve drawn
through the center of the band tends to start from $\beta
A^{1/3}\approx 0$ and terminate at  $\beta A^{1/3}\approx 2$.

\begin{figure}[tb]
\includegraphics[width=0.40\textwidth]{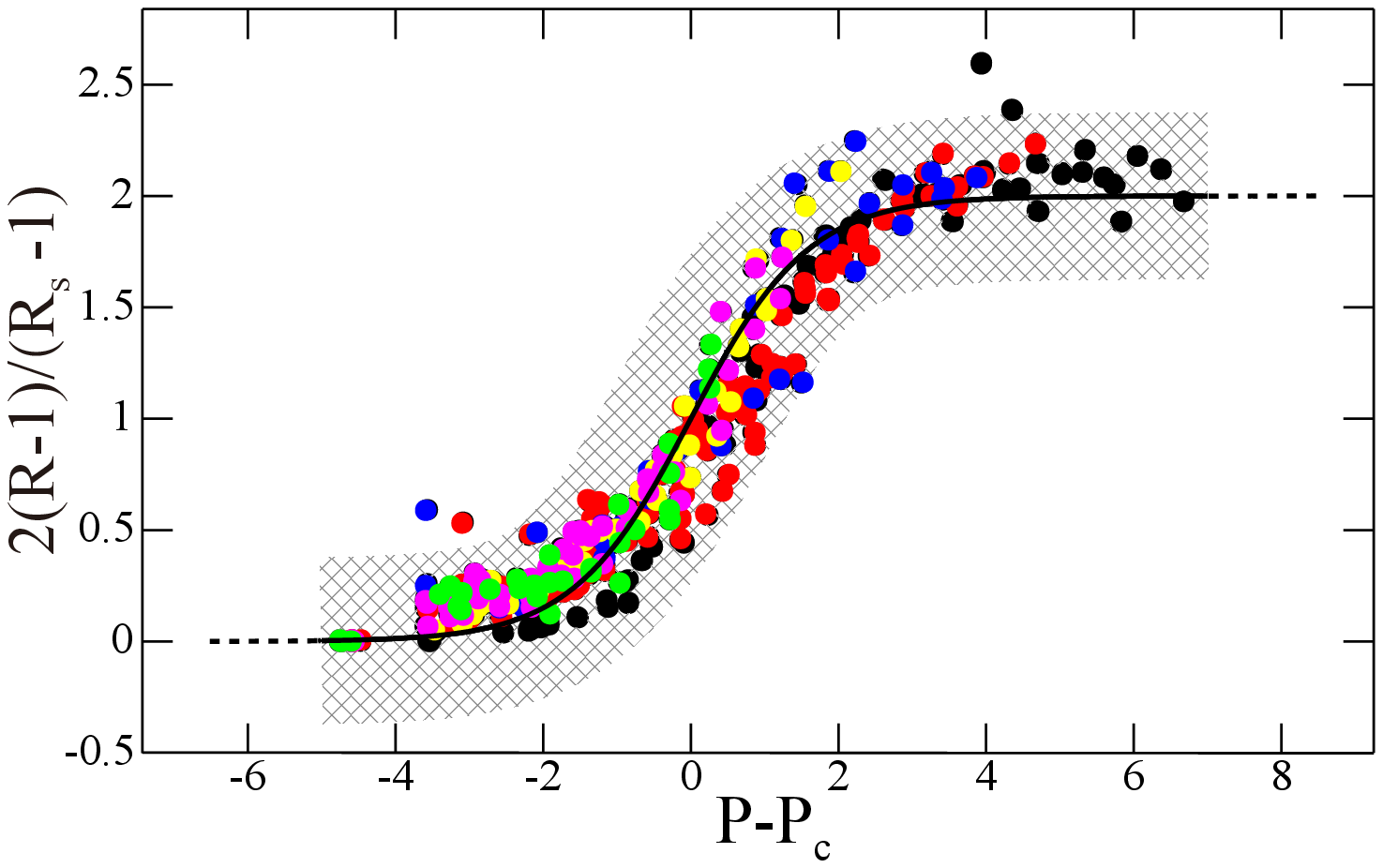}
\caption{\label{fig3}(Color online) Renormalized $2_1^+$ energy
ratio $R$ of the single closed-shell nucleus to its isotope being
studied vs $P-P_c$: Po-Cf (black), Dy-Hg (red), Te-Gd with $N>82$
(blue), Te-Gd with $N<82$ (yellow), Kr-Cd with $N>50$ (pink), and
Ge-Zr with $N<50$ (green), where the saturated energy ratio $R_s$
is taken from Table~\ref{tab1}. The following replacements of
$2_1^+$ energies of single closed-shell nuclei by those in the
nearest available neighbors have been made for the lack of
excitation energies. $^{216}$Th has been tentatively utilized to
replace U-Cf isotopes with $N=126$. We use the $^{134}$Te
($^{98}$Cd) to replace the single closed-shell nuclei of Ce-Hf
(Sr-Mo) isotopes at $N\geq88$ (60). $^{154}$Hf has been used to
replace the single closed-shell nuclei of W-Hg isotopes as
$N_n<16$; $^{206}$Hg has been used to replace the single
closed-shell nuclei of W-Hf isotopes as $N_n\geq16$ (the sudden
change of excitation energy at 16 rather than midshell 22).
$^{128}$Pd has been used to replace $^{126}$Ru for Ru isotopes.}
\end{figure}

A step function with four parameters has been employed to well
reproduce the reduced transition probabilities of actinides in Ref.
\cite{Margraf1992}. The curve drawn in the $\beta A^{1/3}$ $vs$ $(P-P_c)$ plot appears to have a similar form,
but needs only two parameters, $\lambda$ and $P_c$. A universal scaling law is therefore adopted for the
renormalized deformations in the form
\begin{equation}\label{eq1}
\begin{split}
\beta A^{1/3}\approx 1+tanh[\lambda(P-P_c)],
\end{split}
\end{equation}
where $tanh$ is the hyperbolic tangent function, and the constant
$\lambda$~=~0.3135. The function contains two probability
partition factors $e^{\lambda(P-P_c)}$ and $e^{-\lambda(P-P_c)}$
of a Boltzmann-like distribution because the variation of $\beta
A^{1/3}$ in the interval 0 to 2 can be obtained with a
2$e^{\lambda(P-P_c)}/[e^{\lambda(P-P_c)}+e^{-\lambda(P-P_c)}]$
statistical weight. The parameter $\lambda$$P$ is the analog of
$E/kT$ of the Boltzmann distribution. In the limit of an infinite
system, the universal form obviously indicates spherical- and
deformed-shape phases \cite{Cejnar2010,Subir1999}. The two phases
can attain the equilibrium at the inflection point, called
critical point henceforth. Coexistence of deformed and spherical
shapes, as well as fairly similar order parameters, are expected
for nuclei around a given point on the curve.

The most distinctive feature of the scaling law for nuclear shape
is the unification, which reveals a general Boltzmann-like
distribution for even-even nuclei from Germanium to Californium,
emerging from the interplay of the $np$ interaction and
like-nucleon pairing correlations.  Note that it is a
quantum-mechanical behavior essentially different from the
Boltzmann thermodynamic one. It can be shown that the correlation
energy $E$ of $N$ Ising spins \cite{Ising1925} abiding by a
Boltzmann distribution is proportional to $N^2$ (see Eq. 1.12 in
Ref. \cite{Subir1999}). In the case of the nuclear system, there
is also a quadratic dependence of the number of involved nucleons
with the form $ N_nN_p$, which enter in the expression of the
correlation energy. The appearance of the $N_nN_p$ product in the
Boltzmann-like distribution offers the possibility of
reinterpreting its physical significance as a measure of the
correlations embedded in the constituent $\sqrt{N_nN_p}$ effective
$np$ pairs.

For nuclei near the closed shells, the valence neutrons and
protons tend to align anti-parallel to maximize the $np$ residual
interaction \cite{Schiffer1976,Paar1979,Heyde1985} and any two
neighboring $np$ pairs along a chain are expected to be paired to
spin zero along a chain \cite{Bohr1958}. The resultant chain
carries a minimum spin near $J_{np}^{min}=0$. For nuclei near the
mid-shells, parallel orientation is favored in energy for neutrons
and protons in $np$ pairs. This is because the anti-parallel
states are already filled and the interaction in the fully-aligned
configuration is usually the second strongest
\cite{Schiffer1976,Paar1979,Heyde1985}. Furthermore, long-range
correlations dominated by the quadrupole and octupole components
\cite{Dufour1996} exist between any two $np$ pairs along a chain,
which may overcome the short-range pairing correlations occurring
only between the neighboring $np$ pairs and two Ising-like chains.
The long-range correlations therefore result in the parallel
orientation of $np$ pairs. The resultant chain carries a maximal
spin.

In the limit of infinite number of valence nucleons, the angular
momenta of the neighboring neutron and proton along a chain
\cite{Danos1966,Danos1967} are expected to be completely
anti-parallel (lower chains in Fig.~\ref{fig2}) and parallel
(upper chains in Fig.~\ref{fig2}) for spherical- and
deformed-shape phases, respectively. In a general state, $np$
pairs with parallel orientation of the neutron and proton angular
momenta are present in the deformed phase with its statistical
weight
$e^{\lambda(P-P_c)}/[e^{\lambda(P-P_c)}+e^{-\lambda(P-P_c)}]$,
coexisting with $np$ pairs having anti-parallel orientation of the
neutron and proton angular momenta in the spherical
single-particle phase. The deformation of the nucleus in a general
state results from the mixing of the spherical- and deformed-shape
phases, and finally relies only on the $\beta$ of the latter and
its statistical weight, since the deformation $\beta$ of the
spherical-shape phase is zero.

The probability of parallel orientation of the angular momenta of
the neutron and proton in the valence $np$ pairs is small when $P$
is close to zero, reaches up to 0.5 at $P = P_c$, and tends to the
maximum value of 1 in the limit of large $P$. Accordingly, the
deformation undergoes the slow-steep rise up to the critical
point, and thereafter gradually gets saturated. The pairing
correlations show an overall decrease from the lightest to
heaviest nuclei. The pairing correlations lead the $np$ pairs easy
and hard to step into the phase of parallel alignment of the
angular momenta of the neutron and proton for heavy and light
nuclei, respectively. This reveals the significance of the overall
decrease of the critical point from the light to heavy regions in
Table~\ref{tab1}, except for the obvious deviation in the Dy-Hg
region, where the valence nucleons occupy the orbits with the
highest $j$ in the proton and neutron shells, $\pi h_{11/2}$ and
$\nu i_{13/2}$, leading to an enhancement of the $j$-dependent
pairing forces.

After gaining an insight into the orientation of the angular
momenta of the valence neutrons and protons in the $np$ pairs, we
proceed to the two-body correlations between $np$ pairs. The
excitation energy $E_{2^+}$ of the first $2^+$ state as the result
of two-body correlations is another typical example of the
indicator of single-particle or collective degrees of freedom
decreasing along its isotopic chain from the single closed-shell
nucleus to a near-constant value for the mid-shell nuclei. Let $R$
denote the ratio between $E_{2^+}$ of a single closed-shell
nucleus and that of a given isotope. The quantity $R$ varies
regularly from unity to a value $R_s$ close to saturation. The
adopted $R_s$ values for different regions are given in
Table~\ref{tab1}. We define a new parameter $O=(R-1)/(R_s-1)$ and
found that it is universally valid, independent of the mass
region, around the same critical points as the renormalized
deformations. Figure~\ref{fig3} shows that the renormalized energy
ratios are confined in a band similar to that in Fig.~\ref{fig2},
by using the same critical points (a multiplicative factor of 2 in
the numerator is employed to allow the comparison with
Fig.~\ref{fig2}). However, both the rise and saturation are more
rapid. A function similar to that used to fit the renormalized
deformations $\beta A^{1/3}$ has been used to fit the renormalized
energy ratios, in which the exponent in the hyperbolic tangent
function is two times larger.

We initially speculated that the behavior of the parameter $O$ depends on
$e^{2\lambda(P-P_c)}/[e^{\lambda(P-P_c)}+e^{-\lambda(P-P_c)}]^2$,
because $E_{2^+}$ varies inversely proportional to $\beta^2$  \cite{Raman1988},
and $\beta$ has a variation governed by the hyperbolic tangent function involved
in the scaling law of Eq.~\ref{eq1}. The final form obtained from the fit of the experimental data
differs from this general expectation. A possible reason is that we ignore
the contribution of its supplement
$e^{-2\lambda(P-P_c)}/[e^{\lambda(P-P_c)}+e^{-\lambda(P-P_c)}]^2$
present in the spherical-shape phase, leading to a normalized coefficient
$e^{2\lambda(P-P_c)}/[e^{\lambda(P-P_c)}+e^{-\lambda(P-P_c)}]^2 +
e^{-2\lambda(P-P_c)}/[e^{\lambda(P-P_c)}+e^{-\lambda(P-P_c)}]^2$.
Once the $R=1$, $O=0$ normalization point for the spherical-shape phase is adopted,
the final form including $e^{-2\lambda(P-P_c)}$ and
$e^{2\lambda(P-P_c)}$ can be obtained, attesting the
coexistence of two phases in the distribution of the $E_{2^+}$  energies.

In contrast with the critical point, the saturation values $R_s$
of the energy ratio parameter $O$ given in  Table~\ref{tab1} show
a trend which is opposite to that of the critical point parameters
$P_c$. Again, the value for the Dy-Hg region is obviously
abnormal. In order to understand the trend and anomaly, the
analysis of the energy ratio $R$ can help. The $np$ interaction
energy can be considered to cause the decrease of $E_{2^+}$
(namely increase of $R$ until saturation) from a single
closed-shell nucleus towards the mid-shell nuclei
\cite{Shalit1953}. Thus, the increasing trend of the saturated
energy ratio $R_s$ in Table~\ref{tab1} actually reflects the
saturation of the $np$ interaction for high $P$ values in heavy
nuclei, while the opposite decreasing trend of the $P_c$ values as
mentioned above is caused by the pairing correlations.

To summarize, we have shown that hundreds of available nuclear
shapes can be encompassed in a twisted band by fixing the
newly-proposed critical points. The universal form can be well
scaled by a function including the hyperbolic tangent, composed of
two probability partition factors of Boltzmann-like distribution
for spherical- and deformed-shape phases in the limit of an
infinite system, which attain the probability equilibrium at the
critical points. The fraction in the probability partition factors
of Boltzmann-like distribution serving as an analogue to the
Boltzmann exponent manifests itself by the presence of the $np$
interaction between $np$ pairs to like-nucleon pairing
correlations. Excitation energies of the first $2^+$ states
exhibit the same critical points, but quadratic dependence on
partition factors. The Boltzmann-like distribution of nuclear
deformations is further confirmed by the excitation energies of
the first $2^+$ states. We deduce that the partition factors are
related to $np$ anti-parallel and fully-aligned oriented
probabilities responsible for spherical- and deformed-shape
phases, respectively.

\begin{acknowledgments}
This work was supported by the National Natural Science Foundation of China (Grant Nos.~11735017, 11575255, U1732139), the Chinese Academy of Sciences, and the Cai YuanPei 2018 project n. 41458XH.
\end{acknowledgments}

\end{document}